\documentclass[conference,a4paper]{IEEEtran} 

\usepackage{amsmath}
\usepackage{amsfonts}
\usepackage{amssymb}
\usepackage{color}
\usepackage{cite}
\usepackage{graphicx}
\usepackage{epstopdf}

\newcommand*{\diff}{\mathop{}\mathopen{\mathrm{d}}}

\newcommand{\SNR}{\mathsf{SNR}}

\newcommand{\Exp}{\mathsf E}
\newcommand{\expect}[1]{{\Exp}\left[#1\right]}

\newcommand{\markov}{\mathrel\multimap\joinrel\mathrel-%
\mspace{-9mu}\joinrel\mathrel-}

\newcommand{\h}{\mathsf h}
\newcommand{\ent}[1]{{\h}\left(#1\right)}

\newcommand{\I}{I}
\newcommand{\mi}[2]{{\I}\left(#1 \, ; #2 \right)}
\newcommand{\micnd}[3]{{\I}\left(\left. #1 ; #2 \,\right| #3\right)}

\newcommand{\C}{\mathsf C}
\newcommand{\capacity}[1]{{\C}\left(#1 \right)}

\DeclareMathOperator{\erf}{erf}

\title{Upper Bound on the Capacity of Discrete-Time Wiener Phase Noise Channels}
\author{
\IEEEauthorblockN{Luca Barletta and Gerhard Kramer}
\IEEEauthorblockA{Institute for Communications Engineering\\
Technische Universit\"{a}t M\"{u}nchen\\
D-80333 Munich, Germany\\
\{luca.barletta, gerhard.kramer\}@tum.de}}

\begin{document}

\maketitle

\begin{abstract}
 A discrete-time Wiener phase noise channel with an integrate-and-dump multi-sample receiver is studied.
 An upper bound to the capacity with an average input power constraint is derived, and a high signal-to-noise ratio (SNR) analysis is performed.
 If the oversampling factor grows as $\SNR^\alpha$ for \mbox{$0\le \alpha \le 1$}, 
 then the capacity pre-log is at most $(1+\alpha)/2$ at high SNR.
\end{abstract}

\section{Introduction}
\label{Sec:introduction}

Instabilities of the oscillators used for up- and down-conversion of signals in communication systems give rise to the phenomenon known
as \emph{phase noise}~\cite{Demir00}.
The impairment on the system performance can be severe even for high-quality oscillators, if the continuous-time waveform
is processed by long filters at the receiver side.
This is the case, for example, when the symbol time is very long, as happens when using orthogonal frequency division multiplexing.
A study of the signal-to-noise ratio (SNR) penalty induced by filtering of a white phase noise process has been recently 
done in~\cite{BarlettaISIT2014}, where it is shown that the best projection receiver suffers an SNR loss that depends on the
phase noise statistics.

Typically, the phase noise generated by oscillators is a random process with memory, 
and this makes the analysis of the capacity challenging.
The phase noise is usually modeled as a Wiener process, as it turns out to be accurate in describing the phase noise statistics
of certain lasers used in fiber-optic communications~\cite{FoschiniVannucci}, and of free-running microwave oscillators~\cite{Demir00}.
Tight numerical bounds on the information rate of discrete-time phase noise channels with memory are given in~\cite{Barletta2011, Barletta2012,
Barletta2013, BarlettaTIT2014}, while analytical results on single-user Wiener phase noise channels 
are given in~\cite{Lapidoth2002, Barbieri2011, GhozlanISIT2013, GhozlanISIT2014, BarlettaCROWNCOM2014} where it is shown that even weak phase noise
becomes the limiting factor at high SNR.

In~\cite{GhozlanISIT2014} an achievable rate region for the discrete-time Wiener phase noise channel with an integrate-and-dump
oversampling receiver was derived.
For the same channel and receive filter, in this paper we develop an upper bound to the capacity and
characterize the pre-log\footnote{The factor in the capacity high-SNR expansion in front of $\log(\SNR)$.} at high SNR.

The paper is organized as follows. The system model for the continuous-time channel is described in Sec.~\ref{Sec:system}, along with
a simplification that leads to the discrete-time model under consideration. The upper bound to the capacity is derived in Sec.~\ref{Sec:upper},
and the results are discussed in Sec.~\ref{Sec:discussion}. Conclusions are drawn is Sec.~\ref{Sec:conclusion}.

\emph{Notation:} Capital letters denote random variables or random processes. The notation $X_m^n = (X_m,X_{m+1},\ldots, X_n)$ with $n \ge m$
is used for random vectors. With ${\cal N}(0, \sigma^2)$ we denote the probability distribution of a real Gaussian 
random variable with zero mean and variance $\sigma^2$. The symbol $\stackrel{{\cal D}}{=}$ means equality in
distribution.

The symbol $\widetilde{X}$ denotes the reduction of $X$ modulo $[-\pi,\pi)$, and the binary operator $\oplus$ denotes summation modulo $[-\pi,\pi)$. 

Given a complex random variable $X$, we use the notation $X_{||}=|X|$ and $X_{\angle}=\angle X$ 
to denote the amplitude and the phase of $X$, respectively.

The operators $\expect{\cdot}$, $\ent{\cdot}$, and $\mi{\cdot}{\cdot}$ denote expectation, differential entropy, and mutual information,
respectively. 

The function $\log(x)$ denotes the natural logarithm of $x$.

\section{System model}
\label{Sec:system}
In this Section we describe how to obtain a discrete-time version of the continuous-time channel, and we point out the main assumption that leads to the simplified model
analyzed in Sec.~\ref{Sec:upper}.

The output of a continuous-time phase noise channel can be written as
\begin{equation}
 Y(t) = X(t) e^{j\Theta(t)} + W(t),\qquad 0\le t\le T \label{eq:model}
\end{equation}
where $j=\sqrt{-1}$, $X(\cdot)$ is the data-bearing input waveform, and $W(\cdot)$ is a circularly symmetric complex white Gaussian noise.
The phase process is given by
\begin{equation}\label{eq:Theta}
 \Theta(t) = \Theta(0) + \gamma\sqrt{T} B(t/T),\qquad 0\le t\le T,
\end{equation}
where $B(\cdot)$ is a standard Wiener process, \emph{i.e.}, a process characterized by the following properties:
\begin{itemize}
 \item $B(0)=0$,
 \item for any $1\ge t> s\ge 0$, \mbox{$B(t)-B(s)\sim {\cal N}(0,t-s)$} is independent of the sigma algebra generated by \mbox{$\{B(u): u\le s\}$},
 \item $B(\cdot)$ has continuous sample paths almost surely.
\end{itemize}
One can think of the Wiener phase process as an accumulation of white noise:
\begin{equation}\label{eq:ThetaWiener}
 \Theta(t) = \Theta(0) + \gamma \int_0^t B'(\tau)\diff \tau,\qquad 0\le t\le T,
\end{equation}
where $B'(\cdot)$ is a standard white Gaussian noise process.

\subsection{Signals and Signal Space}
\label{subsec:signals}
Suppose $X(\cdot)$ is in the set ${\cal L}^2[0,T]$ of finite-energy signals in the interval $[0,T]$.
Let $\{\phi_{m}(t)\}_{m=1}^\infty$ be an orthonormal basis of ${\cal L}^2[0,T]$. We may write
\begin{align}
X(t)= \sum_{m=1}^\infty X_m \: \phi_m(t),
\quad W(t) = \sum_{m=1}^\infty W_m \: \phi_m(t) \label{eq:W}
\end{align}
where
\begin{align}
X_m &= \int_{0}^{T} X(t) \: \phi_m(t)^\star \diff t,
\end{align}
$x^\star$ is the complex conjugate of $x$, and the $\{W_m\}_{m=1}^\infty$ are independent and identically distributed (iid), 
complex-valued, circularly symmetric, 
Gaussian random variables with zero mean and unit variance.

The projection of the received signal onto the $n-$th basis function is
\begin{align}
 Y_n &= \int_{0}^{T} Y(t) \: \phi_n(t)^\star \diff t\\
 &= \sum_{m=1}^\infty X_m \int_{0}^{T} \phi_m(t) \: \phi_n(t)^\star \: e^{j\Theta(t)} \diff t + W_n \\
 &= \sum_{m=1}^\infty X_m \: \Phi_{mn} + W_n. \label{eq:mimo}
\end{align}
The set of equations given by \eqref{eq:mimo} for $n=1,2,\ldots$ can be interpreted as the output of an infinite-dimensional multiple-input multiple-output channel, whose
fading channel matrix is $\Phi=[\Phi_{mn}]$.

\subsection{Receivers with Finite Time Resolution}
\label{Sec:receiver}
Consider a receiver whose time resolution is limited to $\Delta$ seconds, in the sense that every projection must
include at least a $\Delta$-second interval. More precisely, we set $ML\Delta=T$, where $M$ is the number of independent 
symbols transmitted in $[0,T]$ and $L$ is the oversampling factor, \emph{i.e.}, the number of samples per symbol.
The integrate-and-dump receiver with resolution time $\Delta$ uses the basis functions
\begin{align}
   \phi_m(t) = \left\{ \begin{array}{ll}
   1/\sqrt{\Delta}, & t \in [(m-1)\Delta,m\Delta) \\
   0, & \text{elsewhere}.
   \end{array} \right. \label{eq:rect}
\end{align}
for $m=1,\ldots,ML$.
With the choice~\eqref{eq:rect}, the fading channel matrix $\Phi$ is diagonal and the channel's output for \mbox{$n=1,\ldots,ML$} is
\begin{align}
 Y_n &= X_n \: \frac{1}{\Delta} \int_{(n-1)\Delta}^{n \Delta} e^{j\Theta(t)} \diff t+ W_n \nonumber\\
 &=X_n \: e^{j \Theta((n-1)\Delta)}\frac{1}{\Delta} \int_{(n-1)\Delta}^{n \Delta} e^{j(\Theta(t)-\Theta((n-1)\Delta))} \diff t+ W_n \nonumber\\
 &\stackrel{{\cal D}}{=} X_n \: e^{j \Theta_n}\frac{1}{\Delta} \int_{0}^{\Delta} e^{j\gamma \sqrt{\Delta} B_n(t/\Delta)} \diff t+ W_n \label{eq:Ync}\\
 &\stackrel{(a)}{=} X_n \: e^{j \Theta_n} \int_{0}^{1} e^{j\gamma \sqrt{\Delta} B_n(t)} \diff t+ W_n \nonumber\\
 &=X_n \: e^{j \Theta_n} F_n+ W_n,\label{eq:Yne}
\end{align}
where we have used the notation $\Theta_n=\Theta((n-1)\Delta)$ and $F_n = \int_{0}^{1} e^{j\gamma \sqrt{\Delta} B_n(t)} \diff t$.
In~\eqref{eq:Ync} we have used~\eqref{eq:Theta}, the property $B(t/T)-B((n-1)\Delta/T)\stackrel{{\cal D}}{=}B(t/T-(n-1)\Delta/T)$,
the substitution 
\begin{equation}
 \left\{\begin{array}{l}
	    t\leftarrow t-(n-1)/\Delta \\
	    B_n(t/T) \leftarrow B(t/T - (n-1)\Delta/T),
        \end{array}\right.
\end{equation}
and the property $\sqrt{T}B_n(t/T)\stackrel{{\cal D}}{=} \sqrt{\Delta}B_n(t/\Delta)$. Finally, in step~$(a)$ we have used the
substitution $t \leftarrow t/\Delta$.

Since the oversampling factor is $L$, and the basis functions are square in time domain, we have $X_{kL+1}=X_{kL+2}=\ldots=X_{kL+L}$ for $k=0,\ldots,M-1$, and we can write the model~\eqref{eq:Yne}
as
\begin{equation}\label{eq:contmodel}
 Y_n = X_{\lceil n/L \rceil L} \: e^{j \Theta_n} F_n+ W_n
\end{equation}
for $n=1,\ldots, ML$.

The vectors $X_1^{ML}$, $F_1^{ML}$, and $W_1^{ML}$
are independent of each other. 
The variables $\{X_{kL}\}_{k=1}^M$ are chosen as identically distributed with zero mean and variance
$\expect{|X_{n}|^2}$, and the average power constraint is 
\begin{align}
 \expect{\frac{1}{T}\int_0^T |X(t)|^2 \diff t} &= \frac{1}{ML\Delta}  \sum_{n=1}^{ML} \expect{|X_n|^2 } \nonumber\\
 &=\frac{\expect{|X_n|^2 }}{\Delta}\le {\cal P}. \label{eq:power}
\end{align}
Since we set the power spectral density of $W$ to 1, the power ${\cal P}$ is also the SNR, \emph{i.e.}, 
$\SNR = {\cal P}$.

Using~\eqref{eq:ThetaWiener}, the variables $\Theta_1^{ML}$ follow a discrete-time Wiener process:
\begin{equation}\label{eq:ThetaWiener1}
 \Theta_n = \Theta_{n-1} +  N_{n-1}, \qquad n=1,\ldots, ML,
\end{equation}
where the $N_n$'s are iid Gaussian variables with zero mean and variance $\gamma^2 \Delta$.
The fading variables $F_n$'s are complex-valued and iid, and $F_n$ is independent of $\Theta_1^{n}$.
In other words, $F_n$ is correlated only to $N_n$, and is independent of the vector $(N_1^{n-1},N_{n+1}^{ML})$.

Note that for any finite $\Delta$, or equivalently for any finite oversampling factor $L$, the vector $Y_1^{ML}$ does not
represent a sufficient statistic for the detection of $X$ given $Y$ in the model~\eqref{eq:model}. In other words, the finite time resolution receiver is generally suboptimal.

In this paper we study a simplified model, where the fading variables $F_1^{ML}$ are all one, \emph{i.e.}, we have
\begin{equation}\label{eq:partial}
 Y_n = X_{\lceil n/L \rceil L} \: e^{j \Theta_n} + W_n,\qquad n=1,\ldots, ML.
\end{equation}
This is a commonly-studied model, \emph{e.g.}, see~\cite{GhozlanISIT2014, Martalo2013}, 
and it is referred to as the discrete-time Wiener phase noise channel.
The complete model~\eqref{eq:contmodel} is harder to analyze than the model~\eqref{eq:partial}, because in the former the dependency between
$F_n$ and $N_n$ must be addressed. On the other hand, if the oversampling factor $L$ grows unbounded, then each random variable $F_n$
converges to $1$; this suggests that the analysis of the model~\eqref{eq:partial} can give insights into the analysis of model~\eqref{eq:contmodel}
for receivers with high time resolution.

\section{Upper bound on capacity}\label{Sec:upper}
We compute an upper bound to the capacity of the discrete-time Wiener phase noise channel~\eqref{eq:partial}.
For notational convenience, we use the following indexing for $i=1,\ldots, L$ and $k=1,\ldots, M$:
\begin{equation}\label{eq:model1}
 Y_{(k-1)L+i} = X_k \: e^{j \Theta_{(k-1)L+i}} + W_{(k-1)L+i},
\end{equation}
and we group the output samples associated with $X_k$ in the vector ${\bf Y}_k=Y_{(k-1)L+1}^{(k-1)L+L}$.

The capacity is defined as
\begin{equation}\label{eq:capacity}
 \capacity{\SNR} = \lim_{M\rightarrow \infty}\frac{1}{M}\sup \mi{X_1^M}{{\bf Y}_1^M}
\end{equation}
where the supremum is taken among the distributions of $X_1^M$ such that the average power constraint~\eqref{eq:power} is satisfied.

The mutual information rate can be upper-bounded as follows:
\begin{align}
 \frac{1}{M}&\mi{X_1^M}{{\bf Y}_1^M} = \frac{1}{M}\sum_{k=1}^M\micnd{X_1^M}{{\bf Y}_k}{{\bf Y}_1^{k-1}} \nonumber\\
 & \stackrel{(a)}{\le} \frac{1}{M}\sum_{k=1}^M\micnd{X_1^M}{{\bf Y}_k}{{\bf Y}_1^{k-1}, \widetilde{\Theta}_{(k-1)L}} \nonumber\\ 
 & \stackrel{(b)}{\le} \frac{1}{M}\sum_{k=1}^M\micnd{X_k}{{\bf Y}_k}{\widetilde{\Theta}_{(k-1)L}} \nonumber\\
 & \stackrel{(c)}{=} \micnd{X_1}{{\bf Y}_1}{ \widetilde{\Theta}_{0}} \nonumber\\
 & \stackrel{(d)}{=} \micnd{X_{||,1}}{{\bf Y}_1}{ \widetilde{\Theta}_{0}} + \micnd{X_{\angle,1}}{{\bf Y}_1}{\widetilde{\Theta}_{0},X_{||,1}}   \label{eq:polar}
\end{align}
where step~$(a)$ holds by a data processing inequality and because $\widetilde{\Theta}_{(k-1)L}$ is independent of $X_1^M$, $(b)$ because ${\bf Y}_k$
is conditionally independent of $({\bf Y}_1^{k-1},X_1^{k-1}, X_{k+1}^M)$ given $(X_k,\widetilde{\Theta}_{(k-1)L})$, $(c)$ follows by stationarity
of the processes, and~$(d)$ by polar decomposition of $X_1$ and the chain rule.

For the \emph{amplitude channel}, \emph{i.e.}, the first term in the right-hand side (RHS) of~\eqref{eq:polar}, we have
\begin{align}
 &\micnd{X_{||,1}}{{\bf Y}_1}{ \widetilde{\Theta}_{0}} 
 \stackrel{(a)}{\le} \micnd{X_{||,1}}{ \{X_1 e^{j\widetilde{\Theta}_{i}}+ W_{i}\}_{i=1}^L }{ \widetilde{\Theta}_0^L}  \nonumber\\
 &\qquad\stackrel{(b)}{=} \micnd{X_{||,1}}{ \{X_1 + W_{i}\}_{i=1}^L }{ \widetilde{\Theta}_0^L}  \nonumber\\
 &\qquad\stackrel{(c)}{=}\mi{X_{||,1}}{ \{X_1 + W_{i}\}_{i=1}^L } \nonumber\\
 &\qquad\stackrel{(d)}{=}\mi{X_{||,1}}{ \left|X_1 +\frac{1}{L}\sum_{i=1}^L W_{i}\right| } \nonumber\\
 &\qquad\stackrel{(e)}{\le} \frac{1}{2}\log(1+\SNR) -\frac{\log(2)}{2} + o(1) \label{eq:amplitudechannel}
\end{align}
where~$(a)$ holds by a data processing inequality and because $\widetilde{\Theta}_{0}^L$ is independent of $X_1$, $(b)$ holds due to the circular symmetry 
of the $W_i$'s, $(c)$ because $\widetilde{\Theta}_{0}^L$ is independent of
any other quantity, $(d)$ because the processed variable $\left|X_1 +L^{-1}\sum_{i=1}^L W_{i}\right|$ is a sufficient statistic
for the detection of $X_{||,1}$, and~$(e)$ is an upper bound to the capacity of a non-coherent channel under an average power constraint~\cite[Eq.~(16)]{Lapidoth2002} where $o(1)$
represents a function independent of $L$ that vanishes for $\SNR\rightarrow\infty$.

For the \emph{phase channel}, \emph{i.e.}, the second term on the RHS of~\eqref{eq:polar}, we have
\begin{align}
 &\micnd{X_{\angle,1}}{{\bf Y}_1}{\widetilde{\Theta}_{0},X_{||,1}}  \nonumber \\
 &\qquad= \micnd{X_{\angle,1}}{\{X_1 e^{j\widetilde{\Theta}_{i}}+ W_{i}\}_{i=1}^L}{\widetilde{\Theta}_{0},X_{||,1}} \nonumber \\
 &\qquad\stackrel{(a)}{\le} \micnd{X_{\angle,1}}{\{X_1 e^{j\widetilde{\Theta}_{i}}\}_{i=1}^L}{\widetilde{\Theta}_{0},X_{||,1}} \nonumber \\
 &\qquad\stackrel{(b)}{=} \micnd{X_{\angle,1}}{  \{X_{\angle,1}\oplus \widetilde{\Theta}_{i}\}_{i=1}^L }{\widetilde{\Theta}_{0},X_{||,1}} \nonumber\\
 &\qquad\stackrel{(c)}{\le} \micnd{X_{\angle,1}}{  \{X_{\angle,1}\oplus \widetilde{\Theta}_{i}\}_{i=1}^L }{\widetilde{\Theta}_{0}} \nonumber\\
 &\qquad\stackrel{(d)}{=} \micnd{X_{\angle,1}}{ X_{\angle,1}\oplus\widetilde{N}_{0}, \widetilde{N}_1^{L-1}  }{ \widetilde{\Theta}_{0} } \nonumber\\
 &\qquad\stackrel{(e)}{=} \mi{X_{\angle,1}}{ X_{\angle,1}\oplus\widetilde{N}_{0} } \nonumber\\
 &\qquad=\ent{ X_{\angle,1}\oplus\widetilde{N}_{0} } - \ent{ \widetilde{N}_{0} } \nonumber\\
 &\qquad\stackrel{(f)}{\le}\log(2\pi)- \ent{ \widetilde{N}_{0} } \label{eq:phase1_f}\\
 &\qquad\le \frac{1}{2}\log\left( \frac{2\pi }{e\gamma^2}\right) - \frac{1}{2}\log(\Delta) -g(\Delta)\label{eq:phasechannel}
\end{align}
where in step~$(a)$ we bound by the information extracted by a genie-aided receiver
that knows the additive noise $W_1^L$, $(b)$ is obtained by deleting the amplitude contribution of $X_{||,1}$,
$(c)$ holds because $X_{||,1}\markov (X_{\angle,1},\widetilde{\Theta}_0)\markov \{X_{\angle,1}\oplus \widetilde{\Theta}_{i}\}_{i=1}^L$ forms a Markov chain,
$(d)$ is obtained
by applying reversible transformations, $(e)$
holds because the random variables $(\widetilde{N}_1^{L-1}, \widetilde{\Theta}_{0})$ are independent of any other quantity,
$(f)$ holds by choosing a uniform distribution in $[0,2\pi)$ for $X_{\angle,1}$, 
and the last inequality is derived in the Appendix with
\begin{align}
 g(\Delta) &= \frac{1}{2}\erf\left(\frac{\pi}{\sqrt{2\gamma^2\Delta}}\right)\nonumber\\
 &\quad -\frac{e^{-\frac{\pi^2}{2\gamma^2\Delta}}}{\sqrt{2\pi\gamma^2\Delta}}\left(\pi+\frac{4(\pi+\gamma^2\Delta/\pi)}{1-e^{-\frac{\pi^2}{\gamma^2\Delta}}}\right)-\frac{1}{2}.
\end{align}

Suppose the oversampling factor grows as a power of the SNR, \emph{i.e.}, $L=\Delta^{-1}=\lceil \SNR^\alpha \rceil$ for $0<\alpha\le 1$. 
Inserting~\eqref{eq:amplitudechannel}
and~\eqref{eq:phasechannel} into~\eqref{eq:polar} and using~\eqref{eq:capacity} yields
\begin{align}
 \capacity{\SNR} &\le \frac{1}{2}\log(1+\SNR)-\frac{\log(2)}{2}+\frac{1}{2}\log\left( \frac{2\pi}{e\gamma^2}\right) \nonumber\\
 &\quad+ \frac{1}{2}\log(\lceil \SNR^\alpha \rceil) + o(1),
\end{align}
which for large SNR gives
\begin{align}\label{eq:capacity_upper}
 \limsup\limits_{\SNR\rightarrow\infty} &\left\{ \capacity{\SNR}- \frac{1+\alpha}{2}\log(\SNR) -\frac{1}{2}\log\left( \frac{\pi}{e\gamma^2}\right) \right\} \le 0.
\end{align}

\section{Discussion}\label{Sec:discussion}

As a byproduct of~\eqref{eq:capacity_upper}, an upper bound to the capacity pre-log is
\begin{equation}\label{eq:prelog}
 \lim_{\SNR\rightarrow\infty} \frac{\capacity{\SNR}}{\log(\SNR)} \le \frac{1+\alpha}{2}.
\end{equation}
As shown in the previous Section, a pre-log of $1/2$ comes from the amplitude channel, while a contribution of $\alpha/2$
comes from the phase channel. For example, if no oversampling is used ($L=1$), one can let $\alpha$ go to zero and obtain 
just the degrees of freedom provided by the amplitude channel, \emph{i.e.}, $1/2$. This means that, without oversampling,
the Wiener phase noise channel has the same degrees of freedom of the non-coherent channel. This is in accordance with the result
given in~\cite{Lapidoth2002}.

If the oversampling factor grows as $\sqrt{\SNR}$, for $\alpha=1/2$, then a pre-log higher than $3/4$ can not be achieved. Indeed, a
pre-log as high as $3/4$ can be achieved with the processing described in~\cite{GhozlanISIT2014}: the amplitude channel 
contributes with pre-log $1/2$ by using the statistic
\begin{equation}
 V_k = \sum_{i=1}^L |Y_{(k-1)L+i}|^2
\end{equation}
to detect $X_{k,||}$, and the phase channel contributes with pre-log $1/4$ by using the processing
\begin{equation}
 \angle \widetilde{Y}_k = \angle \left( Y_{(k-1)L+1} \left(\frac{Y_{(k-1)L}}{X_{k-1}}\right)^\star \right)
\end{equation}
to detect $X_{k,\angle}$.

\begin{figure}
 \centering
 \includegraphics[width=\columnwidth]{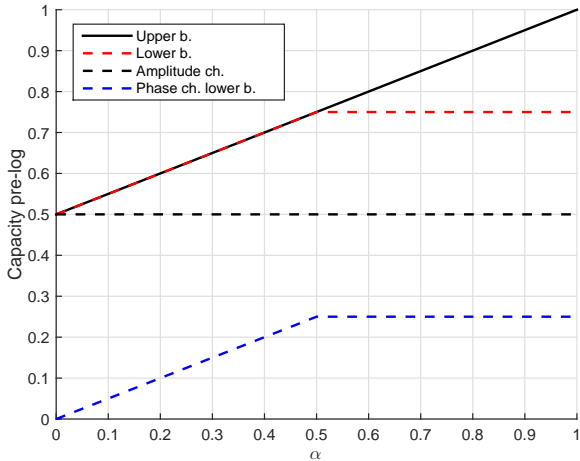}
 \caption{Capacity pre-log bounds as a function of $\alpha$ for high SNR. The oversampling
 factor $L$ is $L=\lceil \SNR^\alpha \rceil$.}
 \label{fig:prelog}
\end{figure}
Figure~\ref{fig:prelog} plots the known upper and lower bounds to the capacity pre-log 
at high SNR. The upper bound is the result of this paper, expressed in~\eqref{eq:prelog}. The lower bound is based on results 
derived in~\cite{GhozlanISIT2014}. More specifically:
\begin{itemize}
 \item the lower bound for the amplitude channel, shown as the dashed black line, was derived independent
  of the growth rate of the oversampling factor;
 \item for the phase channel it was shown how to achieve pre-log $1/4$ for $\alpha = 1/2$, hence the same pre-log can be achieved for 
  $1/2\le \alpha \le 1$. It is not difficult to use the results of~\cite{GhozlanISIT2014} to extend the lower bound 
  in the range $0\le \alpha<1/2$. It turns out that the achievable pre-log linearly increases from $0$ to $1/4$.
\end{itemize}
From the Figure, the capacity pre-log is exactly known in the range $0\le \alpha \le 1/2$, where the upper and lower
bounds agree. 
The upper bound derived in this paper does not rule out the possibility to achieve pre-log $1$ if the oversampling
factor grows faster than $\sqrt{\SNR}$.

Consider the case of receivers without oversampling ($L=1$ or $\alpha\rightarrow 0$). The analysis of Sec.~\ref{Sec:upper} shows that the simplified model~\eqref{eq:partial} has capacity pre-log $1/2$, while the general 
discrete-time model that accounts also for the amplitude fading $F_n$~\eqref{eq:contmodel} has a $\log(\log(\SNR))$ behavior at high SNR~\cite{Lapidoth2003}, \emph{i.e.}, pre-log $0$.
This means that, at least in the case $L=1$, the simplified model is not a good approximation of the complete model.

\section{Conclusions}\label{Sec:conclusion}

We have derived an upper bound to the capacity of discrete-time Wiener phase noise channels.
As a byproduct, we have obtained an upper bound to the capacity pre-log at high SNR that depends on the growth rate
of the oversampling factor used at the receiver. If the oversampling factor grows proportionally to $\SNR^\alpha$,
then a capacity pre-log higher than $(1+\alpha)/2$ can not be achieved.

Previous results on a lower bound to the capacity pre-log allow to state that the capacity at high SNR is exactly
$(1+\alpha)/2\: \log(\SNR)$ for $0\le \alpha \le 1/2$.

\appendix
\section*{A lower bound to $\ent{ \widetilde{N}_0 }$}

The probability density function of $\widetilde{N}_0$ is
\begin{equation}
 p_{\widetilde{N}_0}(x) = \sum_{k=-\infty}^\infty p_{N_0}(x+2\pi k) = \sum_{k=-\infty}^\infty \frac{1}{\sqrt{2\pi \gamma^2 \Delta}} e^{-\frac{(x+2\pi k)^2}{2\gamma^2\Delta}}
\end{equation}
for $-\pi\le x<\pi$ and zero elsewhere, and can be upper-bounded as follows for $-\pi \le x < \pi$:
\begin{align}
 p_{\widetilde{N}_0}(x) &\stackrel{(a)}{\le} \frac{1}{\sqrt{2\pi \gamma^2 \Delta}} \left(e^{-\frac{x^2}{2\gamma^2\Delta}}+2\sum_{k=0}^\infty  e^{-\frac{\pi^2(2k+1)^2}{2\gamma^2\Delta}} \right)\nonumber\\
 &\stackrel{(b)}{\le} \frac{1}{\sqrt{2\pi \gamma^2 \Delta}} \left(e^{-\frac{x^2}{2\gamma^2\Delta}}+2\sum_{k=0}^\infty  e^{-\frac{\pi^2(2k+1)}{2\gamma^2\Delta}} \right)\nonumber\\
 &= \frac{1}{\sqrt{2\pi \gamma^2 \Delta}} \left(e^{-\frac{x^2}{2\gamma^2\Delta}}+2\frac{e^{-\frac{\pi^2}{2\gamma^2\Delta}}}{1-e^{-\frac{\pi^2}{\gamma^2\Delta}}} \right) \label{eq:app_pdf}
\end{align}
where step~$(a)$ follows by using $x\ge -\pi$ for the terms with $k\ge 1$ and $x< \pi$ for the terms with $k\le -1$.
Inequality~$(b)$ holds because $(2k+1)^2\ge 2k+1$ for $k\ge 0$.
The differential entropy of $\widetilde{N}_0$ can be lower-bounded as follows:
\begin{align}
 &\ent{\widetilde{N}_0} = \expect{- \log \left(p_{\widetilde{N}_0}(\widetilde{N}_0)\right) } \nonumber\\
 &\stackrel{(a)}{\ge} \frac{1}{2}\log(2\pi \gamma^2 \Delta)+\frac{\expect{\widetilde{N}_0^2}}{2\gamma^2\Delta}-\expect{\log\left(1+\frac{2 e^{\frac{\widetilde{N}_0^2-\pi^2}{2\gamma^2\Delta}}}{1-e^{-\frac{\pi^2}{\gamma^2\Delta}}}\right)}\nonumber\\
 &\ge \frac{1}{2}\log(2\pi \gamma^2 \Delta)+\frac{\expect{\widetilde{N}_0^2}}{2\gamma^2\Delta}-\frac{2 e^{-\frac{\pi^2}{2\gamma^2\Delta}}}{1-e^{-\frac{\pi^2}{\gamma^2\Delta}}}\expect{e^{\frac{\widetilde{N}_0^2}{2\gamma^2\Delta}}}\label{eq:app_ent1}
\end{align}
where inequality~$(a)$ is due to~\eqref{eq:app_pdf} and the monotonicity of the logarithm, and the last inequality is due to $\log(1+x)\le x$. A lower bound to the second moment of $\widetilde{N}_0$ is:
\begin{align}
 \expect{\widetilde{N}_0^2} &= \sum_{k=-\infty}^\infty \int_{-\pi}^\pi x^2 p_{N_0}(x+2\pi k) \diff x \nonumber\\
 &\ge \int_{-\pi}^\pi x^2 p_{N_0}(x) \diff x \nonumber\\
 &=  \gamma^2\Delta \erf\left(\frac{\pi}{\sqrt{2\gamma^2\Delta}}\right) - \sqrt{2\pi\gamma^2\Delta} e^{-\frac{\pi^2}{2\gamma^2\Delta}} \label{eq:app_2ndmoment}
\end{align}
where 
\begin{equation}
 \erf(x) = \frac{2}{\sqrt{\pi}} \int_0^x e^{-t^2} \diff t
\end{equation}
is the error function. Since all the terms of the summation are positive, the inequality follows by considering only the term for $k=0$.
An upper bound to the last expectation on the RHS of~\eqref{eq:app_ent1} is
\begin{align}
 &\expect{e^{\frac{\widetilde{N}_0^2}{2\gamma^2\Delta}}} = \frac{1}{\sqrt{2\pi \gamma^2 \Delta}}\sum_{k=-\infty}^\infty \int_{-\pi}^\pi  e^{-\frac{(x+2\pi k)^2-x^2}{2\gamma^2\Delta}} \diff x \nonumber\\
 &= \frac{1}{\sqrt{2\pi \gamma^2 \Delta}}\sum_{k=-\infty}^\infty \frac{\gamma^2\Delta}{ \pi k} e^{-\frac{2 \pi^2 k^2}{\gamma^2\Delta}} \sinh\left(\frac{2\pi^2 k}{\gamma^2\Delta}\right) \nonumber\\
 &= \frac{2}{\sqrt{2\pi \gamma^2 \Delta}}\left(\pi+\sum_{k=1}^\infty \frac{\gamma^2\Delta}{\pi k} e^{-\frac{2 \pi^2 k^2}{\gamma^2\Delta}} \sinh\left(\frac{2\pi^2 k}{\gamma^2\Delta}\right)\right) \nonumber\\
 &\le \frac{2}{\sqrt{2\pi \gamma^2 \Delta}}\left(\pi+\sum_{k=1}^\infty \frac{\gamma^2\Delta}{\pi } e^{-\frac{2 \pi^2 k^2}{\gamma^2\Delta}} \sinh\left(\frac{2\pi^2 k}{\gamma^2\Delta}\right)\right) \nonumber\\
 &= \frac{2}{\sqrt{2\pi \gamma^2 \Delta}}\left(\pi+ \frac{\gamma^2\Delta}{\pi } \right).\label{eq:app_expmoment}
\end{align}
The bound~\eqref{eq:app_ent1} with inequalities~\eqref{eq:app_2ndmoment} and~\eqref{eq:app_expmoment} give
\begin{align}
 \ent{\widetilde{N}_0} &\ge \frac{1}{2}\log(2\pi e \gamma^2 \Delta)+\frac{1}{2}\erf\left(\frac{\pi}{\sqrt{2\gamma^2\Delta}}\right)\nonumber\\
 &\quad  -\frac{e^{-\frac{\pi^2}{2\gamma^2\Delta}}}{\sqrt{2\pi\gamma^2\Delta}}\left(\pi+\frac{4(\pi+\gamma^2\Delta/\pi)}{1-e^{-\frac{\pi^2}{\gamma^2\Delta}}}\right)-\frac{1}{2}\nonumber\\
 &=\frac{1}{2}\log(2\pi e\gamma^2 \Delta)+g(\Delta).\label{eq:app_result}
\end{align}
Also, note that $\ent{\widetilde{N}_0}\le \ent{N_0}=1/2\cdot\log(2\pi e \gamma^2\Delta)$, so we have that the bound~\eqref{eq:app_result} is tight for small $\Delta$, \emph{i.e.},
\begin{align}
 \lim_{\Delta\rightarrow 0} g(\Delta) = 0.
\end{align}

\section*{Acknowledgment}
L. Barletta was supported by the Technische Universit\"{a}t M\"{u}nchen –- Institute for Advanced Study, funded by the German Excellence Initiative.
G. Kramer was supported by
an Alexander von Humboldt Professorship endowed by the
German Federal Ministry of Education and Research.

\bibliographystyle{IEEEtran}
\bibliography{refSuffStat}

\end{document}